\documentstyle[prl,aps,preprint]{revtex}
\begin{document}
\draft

\title{\bf
de Haas-van Alphen Oscillations in a Superconducting State at High Magnetic
Fields}
\author{
 Sa\v{s}a Dukan\thanks{On leave from Rugjer Bo\v{s}kovi\' c Institute, P.O.
Box 1016, Zagreb, Croatia}
 and Zlatko Te\v {s}anovi\' {c}}
\address{\sl
Department of Physics and Astronomy,
Johns Hopkins University,
Baltimore, MD 21218, USA}
\maketitle
\begin{abstract}
{}~~Low-temperature quantum oscillations of the dHvA amplitude are shown
to persist far below the upper critical field
of a strongly type-II superconductor, due to the gapless nature of
the BCS quasiparticle spectrum in high fields.
The dHvA amplitude in the superconducting state is smaller
than its normal state counterpart
by the factor $\sim [max(T,\Gamma )/\Delta ]^2$, where $\Gamma$
is the damping.  This factor reflects the presence of a small
gapless portion of the Fermi surface, surrounded
by regions where the BCS gap is large.
The agreement with recent experimental data on $V_3Si$ is very good.
\end{abstract}
\narrowtext

There has been much interest lately in
properties of high T$_c$ superconductors (HTS) and
other strongly type-II systems at high magnetic fields\cite{osc}.
Recent reports clearly demonstrate
the de Haas-van Alphen effect
in the mixed state of A15 superconductors ($V_3Si$ and $Nb_3Sn$), as well as
in the layered superconductor $2H-NbSe_2$ \cite{hayden,corco,gr}. In all cases
it was found that
quantum oscillations in the dHvA amplitude persist
to a surprisingly low fraction ($\sim$ 60\%) of
the upper critical field $H_{c2}$. This is surprising since,
in the standard Abrikosov-Gor'kov theory, one expects an
exponential suppression of dHvA amplitude due to the
large superconducting gap $\Delta$
at the Fermi surface. Furthermore, it
was found that the dHvA amplitude for the
fixed value of magnetic field, $H$, behaves as
a function of temperature, $T$, in the same way as in
the normal state, except for the overall reduction
in magnitude when the sample
becomes superconducting. The presence of these dHvA
oscillations in the
mixed phase could, in principle,
be due to a small portion of the sample
remaining normal. However, heat
capacity measurements have ruled out this
possibility\cite{hayden,corco}.

In this Letter we show that these experimental results are a
direct manifestation of a qualitatively new nature
of the BCS quasiparticle spectrum at high fields.
As shown in Refs. \cite{sasa,sasa2}, at fields near $H_{c2}$,
this spectrum is gapless at the set of points in the magnetic
Brillouin zone (MBZ). These nodes in the gap reflect the
{\it center-of-mass} motion of Cooper pairs in high magnetic field,
in contrast to the familiar nodes of a zero-field
unconventional [i.e., p- or d-wave] anisotropic superconductor
which are due to the {\it relative}
orbital motion.  The gapless
behavior persists to a relatively low fraction of $H_{c2}$
\cite{sasa2}, as long as $\Delta (T,H)$ remains smaller
than or comparable to $\hbar\omega_c$,
where $\Delta (T,H)$ is
the average BCS gap and $\omega_c=eH/mc$ is the
cyclotron frequency.  The results of
Corcoran {\it et al.} then follow
from the presence of a small portion of the Fermi surface
containing a coherent gapless band of quasiparticles, while
the rest of it is gapped by a large $\Delta$ \cite{footi}.
High magnetic fields, low temperatures and clean samples
provide ideal conditions for the validity of this
picture, with the coherent
quasiparticle propagation extending over many
unit cells of the vortex lattice.
For example, the $V_3Si$ sample used in experiment
\cite{hayden} satisfies well the last condition,
with its  electronic
mean-free path, $l_0$, being much longer than
the separation between the vortices (given
by the magnetic length $l\equiv \sqrt{c/2eH}$):
$l_0=1450$ \AA  ~at $T=1.3K$,  while the
intervortex separation varies
from $\sim 60 $\AA  ~to $\sim 45$ \AA  ~for fields between $10$
and $18.5$ Tesla.
This situation should be contrasted with the one
at low fields, where coherent propagation is
suppressed and the low-lying quasiparticle excitations
are localized in the cores of isolated vortices.

At high magnetic fields, the electrons
participating in superconducting pairing occupy
a well-defined set of Landau levels (LLs). At low temperatures
and high fields, where
$\hbar \omega_c\gg T,\tau ^{-1}$
($\Gamma_0=\hbar/2\tau$ is the scattering rate) and
$\hbar\omega_c\agt\Delta$, numerous
`quantum oscillation' effects
arise as consequence of this Landau
quantization\cite{quantum}.  In this regime
the superconducting order parameter $\Delta({\bf r})$
is well-described by the Abrikosov solution \cite{abr}
confined to the lowest LL
for Cooper pairs\cite{zbt}.
The BCS Hamiltonian is diagonalized in
the representation of Ref. \cite{bychkov},
where electronic states are labeled
by the quasi-momentum ${\bf q}$ [${\bf q}\perp {\bf H}$]
restricted to the first MBZ, momentum
$k_z$ parallel to the field and the LL index $n$.
Cooper pairs are
formed from the electrons with
the opposite momenta $({\bf q},k_z)$ and spins
belonging to the same (diagonal pairing) or different LLs
(off-diagonal
pairing)\cite{sasa,zbt}.
Near $H_{c2}$, where $\Delta \ll \hbar \omega _c$,
the off-diagonal pairing can be treated as a perturbation and
the BCS quasiparticle spectrum
is obtained analytically
\cite{sasa,sasa2,zbt}:
\begin{eqnarray}
E_{k_{z},{\bf q},n}
=\pm \sqrt{\varepsilon _{n}(k_z)^2 + |\Delta_{nn}({\bf q})|^{2}}
\label{spectrum}
\\
\nonumber
\varepsilon _{n}(k_z)=\frac{\hbar ^2k_z^2}{2m}+\hbar \omega _c(n+1/2)-\mu ,
\end{eqnarray}
where $\mu$ is the chemical potential.
The spectrum consist of $n_c$
branches ($n_c={\rm int}[\mu /\hbar
\omega _c]$ is the number of occupied Landau
levels) in which the superconducting
order parameter $\Delta _{nn}({\bf q})$
goes to zero at points ${\bf q_j}$, forming a `vortex lattice' in the
${\bf q}$-space. There is a strong
linear dispersion around these points (nodes).
$\Delta _{nn}({\bf q})$ for different $n$ in (\ref{spectrum})
behave similarly in the vicinity of
and differ considerably only far away from these nodes.
At lower fields,
where $\Delta$ becomes comparable to $\hbar\omega _c$, the off-diagonal
pairing must be included on equal footing. The excitation
spectrum is found numerically
in Ref. \cite{sasa2} and still exhibits
gapless behavior in some of the
branches, while the gap starts
opening up in the rest. Eventually,
at yet lower fields, the LL structure is destroyed
by large $\Delta$.\cite{sasa2}  The presence of these nodes
and other regions with a very small gap at the Fermi surface,
as $\Delta (T,H)$ increases from zero to $\agt\hbar\omega_c$,
is the key aspect of our theory.

We now consider the dHvA effect in such gapless high-field
superconductor in the low temperature limit.
The first step is to determine the {\sl oscillatory}
part of the thermodynamic potential, $\Omega _{osc}(H,T)$.
Initially, we consider a homogeneous
3-D system [like $V_3Si$, $Nb_3Sn$ and
other A15 type-II superconductors] and then show how the results
change for the layered systems like $NbSe_2$.

The quasiparticle contribution to the
thermodynamic potential per unit volume is
given by:
\begin{equation}
\Omega _{exc}= -\frac{1}{\beta }\sum _{spin}
\frac{1}{L_xL_yL_z}\sum_{n=0}^{n_c}\sum_{k_z{\bf q}}
\left\{\ln \left[1+\exp (-\beta E_n(k_z,{\bf q}))\right]
+\frac{\beta }{2}\left[E_n(k_z,{\bf q})-\varepsilon _n(k_z)\right]\right\}
\label{omega}
\end{equation}
where $\beta =1/T$
and  $E_n(k_z,{\bf q})$ is the quasiparticle excitation
energy (\ref{spectrum}).
Using the standard Poisson resummation formula:
\begin{equation}
\sum_{n=0}^{\infty }\phi (n)=\phi (0)/2+\int_{0}^{\infty }dn\phi (n)+2{\sl Re}
\sum_{k=1}^{\infty }\int_{0}^{\infty }dn\phi (n)\exp(i2\pi kn)
\label{poisson}
\end{equation}
we can perform the sum over Landau level index $n$ in (\ref{omega}) and get
the oscillatory part of $\Omega _{exc}$ as:
\begin{eqnarray}
\Omega _{osc}=-\frac{1}{2\pi ^3l^3}\frac{1}{\beta \hbar \omega _c}\sum_{k=1}^
{\infty }\frac{(-1)^{k}}{\sqrt{k}}\cos{(\frac{2\pi k}{\hbar \omega _c}
\mu-\frac{\pi}{4})}\int_{-\infty }^{\infty }d\varepsilon
\exp{(i2\pi k\varepsilon /\hbar
\omega _c)}\\
\label{osc}
\int_{MBZ}d^2\tilde{q}\left\{\ln\left[1+\exp(-\beta
\sqrt{\varepsilon ^2+|\Delta ({\bf \tilde{q}})|^2})\right]
+\frac{\beta }{2}\left[\sqrt{\varepsilon ^2+|\Delta ({\bf \tilde{q}})|^2}-
\varepsilon \right]\right\}
\nonumber
\end{eqnarray}
where $\tilde{q}=ql$ is the quasi-momentum
rescaled by magnetic length.

In deriving (4) we have
assumed a spherical Fermi surface. In real
systems, with more complicated
Fermi surfaces, $\cos{(2\pi k \mu /
\hbar \omega_c-\pi /4)}$ is replaced
by $\cos{(\hbar A/\omega _{c}^*m^*-\pi /4)}$,
where $A$ is the extremal cross-sectional area of the Fermi surface in
the plane
$\perp {\bf H}$ and $\omega _{c}^*=eH/m^*c$, $m^*$
being the effective mass associated with the orbit around $A$
\cite{lif}). We have also neglected the fact that
$\Delta $, the overall amplitude of
$\Delta _{nn}({\bf q})$,
oscillates as a function of $H$,
and have assumed that the oscillatory
piece of $\Delta $ is much smaller then its `smooth' part.
This is a good approximation
if the number of occupied LLs,
$n_c$, is large [$\sqrt{n_c}\gg 1$].
Furthermore, we have neglected
different features in $\Delta_{nn}({\bf q})$ as a function of a LL
index $n$ since they are pronounced
only away from the gapless points [nodes].

While the systems in question
are rather clean\cite{corco}, some impurities and defects are still present
and give rise to a small exponential decay of dHvA amplitude
even in the normal state. Therefore, we should include disorder
in our calculations. This is accomplished
by introducing the density of states per gapless branch of the spectrum
(\ref{spectrum}) in presence of disorder as:
\begin{equation}
N(\omega )/N_n(0)=
\frac{1}{\pi }\Im m \sum_{{\bf q},k_z}
\frac{\tilde{\omega }+\varepsilon (k_z)}
{-\tilde{\omega }^2+E({\bf q},k_z)^2}~~,
\label{density}
\end{equation}
where $\tilde{\omega }=\omega -\Sigma (\tilde{\omega })$ and $N_n(0)$ is the
normal density of states at the LL. In the clean limit
(no disorder) $N(\omega )/N_n(0)\approx (\omega /\Delta )^2$ for small
energies,
reflecting the presence of nodes (gapless points) in the quasiparticle
spectrum. This strong energy dependence suggests
that the effective scattering rate $\tau ^{-1}$ in the
superconducting
state will be modified relative to the
normal scattering rate $\tau _o^{-1}$.
We take the self-energy $\Sigma (\tilde{\omega })$ to be $\Sigma (\tilde{
\omega })=-i\Gamma$,
where $\Gamma =\hbar /2\tau $, which results in $N_{dis}(\omega )/N_n(0)
\approx (\Gamma /\Delta )^2+(\omega /\Delta )^2$.
This is a good approximation in the
{\sl unitary} limit (otherwise the
scattering rate depends strongly on energy--we
have investigated the behavior of the density
of states in presence of various forms of static disorder
and will present these results elsewhere
\cite{sasa3}). In the end, $\Gamma $ has to be
determined self-consistently as $\lim _{\omega
\rightarrow 0}i\Sigma (\tilde{\omega })$.

It is known that
in conventional
superconductors, with a finite
gap everywhere in the momentum space,
the quasiparticle contribution to
$\Omega _{osc}(H,T)$
is exponentially small at low
temperatures because of the large value of
$\Delta/T$ (unless the field is very close
to $H_{c2}$, within few percent). On the other hand, in the
gapless high-field superconductor
described above, there are quasiparticle
excitations with momenta ${\bf q}$ such that
$\Delta ({\bf q})\leq T$ which give a large contribution to  the
thermodynamic potential (4). Therefore, we divide the
MBZ in two different regions. The `gapless' region
(which we label $\cal G$),
has $\Delta ({\bf \tilde{q}})\cong\Delta \tilde{q}$ for
all $\tilde{q}\leq \tilde{q_c}$, with $\tilde{q_c}\cong
C max(T/\Delta ,\Gamma /\Delta )$ being the radius of $\cal G$
and $C$ a constant of order unity.
Outside of $\cal G$, where
$\Delta ({\bf \tilde{q}})>max(T,\Gamma )$ for
all $\tilde{q}$, we will aproximate
$\Delta ({\bf \tilde{q}})$ with the average value of the gap function
${\Delta}=<\Delta ({\bf \tilde{q}})>_{av}$
\cite{stephen}. This is the `gapped' region.
In setting the boundaries of $\cal G$ we have
allowed the possibility of $T<\Gamma $, which is
actually the case in
Ref. \cite{hayden}. There, for example, $T<0.2$ meV while
the normal scattering rate for
$V_{3}Si$ superconductor is $\Gamma _0=0.36$ meV in
the normal state (we will show that
the self-consistent $\Gamma $ actually increases slightly
from the normal state value).

Inserting the density of states
(\ref{density}) in (4) and taking the
derivative over $H$, we obtain the oscillatory
part of the magnetization due
to the qasiparticle excitations from the region $\cal G$ as:
\begin{eqnarray}
M_{osc}^{{\cal G}}(H,T)=-\frac{e\hbar }{mc}\frac{1}{\pi l^3}
\frac{\mu }{(\hbar \omega _c)^2}
\left[C max(\frac{T}{\Delta },\frac{\Gamma }{\Delta })\right] ^2
\sum_{k=1}^{\infty }\frac{(-1)^k}{\sqrt{k}}\sin{(\frac
{2\pi k\mu}{\hbar \omega_{c}}-\pi /4)}
\\
\label{magG}
\nonumber
\times T/\sinh{(\frac{2\pi ^2
kT}{\hbar \omega _c})}\exp{(-2\pi k\Gamma /\hbar \omega _c)}
+{\cal O}\left( [max(\frac{T}{\Delta },\frac{\Gamma }{
\Delta })]^4\right)
\end{eqnarray}
The amplitude of dHvA oscillations ${\cal A}(H,T)$
of magnetization (6) for the
first Fourier harmonic k=1 is:
\begin{equation}
{\cal A}^{{\cal G}}(H,T)\propto 2\left[C max(\frac{T}{\Delta },
\frac{\Gamma }{\Delta })\right] ^2
 H^{-1/2}\frac{T}{\sinh{(2\pi ^2T/\hbar \omega _c)}}
\exp{(-2\pi \Gamma /\hbar \omega _c)}
\label{alpha}
\end{equation}
which is the same contribution as in the normal state reduced by the factor
$G=2[C max(T/\Delta ,\Gamma /\Delta )]^2$ (for the moment we assume
$\Gamma \sim \Gamma _0$). Factor $2$ in (\ref{alpha}) comes
from the presence of two gapless points in the First MBZ of the spectrum
(\ref{spectrum}). The result (\ref{alpha}) tells us that,
in passing from the normal to the
superconducting state, there should be a drop
in the magnitude of dHvA oscillations,
since now there is only a small fraction $G\ll 1$
of the Fermi surface where there are low-energy excitations.
The size of ${\cal G}$
is determined by both the total number of nodes at the
Fermi surface and the areas in different branches where
the BCS gap is very small although not necessarily zero.
In presence of finite disorder both will broaden
into gapless regions \cite{footx}.  At low temperatures,
such that $T< \Gamma $, this broadening is
quite small: only $\sim$3\% of Fermi surface
in $V_3Si$ is still gapless at $T=1.3$K
by the time magnetic field
drops from $H=18.5 (\sim H_{c2})$ Tesla to $H=10$ Tesla.

The quasiparticle excitations
with momenta outside $\cal G$,
where $\Delta >max(T,\Gamma )$ ,
contribute to the amplitude of dHvA oscillations as:
\begin{equation}
{\cal A}^{1-{\cal G}}(H,T)\propto (1-G)H^{-1/2}\frac{\hbar \omega _c}{2\pi ^2}
f(\frac{2\pi \Delta }{\hbar \omega _c})\exp{(-2\pi \Delta /\hbar \omega _c
-2\pi \Gamma /\hbar \omega _c})
\label{extra}
\end{equation}
where $f(x)=1$ for $x=0$ and $f(x)=\sqrt{\pi x/2}$ for $x\gg 1$ and we have
assumed $k=1$. This amplitude is much smaller then ${\cal A}^{\cal G}$: it
is already $\sim 20$ times smaller then (7) at $H=16.5$ Tesla in $V_3Si$
system and becomes negligible at lower fields.

Figure 1 shows the field dependence
of $ln[{\cal A}(H,T)\sinh{(2\pi ^2T/\hbar
\omega _c)}H^{1/2}T^{-1}]$
[Dingle plot] for $V_3Si$ sample used in the experiment by
Corcoran {\it et al.}\cite{hayden},
both in normal and superconducting states.
The normal state data (straight line) give the normal scattering
rate of $\Gamma _0=0.36$ meV. The dashed line
in Figure 1 shows the theoretical plot obtained with
${\cal A}(H,T)={\cal A}^{\cal G}(H,T)+{\cal A}^{1-{\cal G}}(H,T)$,
where we have taken
the scattering rate $\Gamma $ equal to
its normal value $\Gamma _0$.
We have assumed  $\Delta (H)=\Delta (0)\sqrt{1-H/H_{c2}}$,
which is a good approximation for the range of fields used
in the experiment. The comparison of this curve and
the experimental points suggests that there
is a small increase in the
scattering rate in the mixed phase.
Indeed, the self-consistent calculation
of $\Gamma $ gives \cite{footx}:
\begin{equation}
\Gamma (H)=\sqrt{\frac{\Gamma _{0}\Delta (H)}{2}}~~~~~
{\rm for}~~~~~~\Gamma /\Delta <1
\label{gamma}
\end{equation}
Using this value of $\Gamma$ in (7) and (8)
we obtain the full line in Figure 1. Our theoretical curve
is in excellent agreement with the experimental data. We
should note here that we use only
a single fitting parameter, $C\sim 1.012$,
which yields a reasonable value of $\sim$
3\% for the size of ${\cal G}$.
All other quantities are taken from Ref. \cite{corco}.

Finally, we address the problem  of dHvA oscillations in layered
superconductors, like $2H-NbSe_2$ and HTS
$YBa_2Cu_3O_{7-\delta }$. The layered systems
have been extensively studied ever
since Graebner and Robbins \cite{gr} reported
first observation of dHvA
oscillations in $2H-NbSe_2$ at
magnetic fields below $H_{c2}$.
In a layered system there is hopping, $t$,  between
the layers so that the electron dispersion along the
field is given by $t\cos{(k_zd)}$,
where $d$ is the interlayer separation.
In this case the oscillatory part of the
magnetization due to the quasiparticle
excitations around gapless points is given by:
\begin{eqnarray}
M_{osc}^{{\cal G}}(H,T)=-\frac{e\hbar }{mc}\frac{1}{l^2d}\frac{\mu }
{(\hbar \omega _{c})^{3/2}t^{1/2}}
\left[C max(\frac{T}{\Delta },\frac{\Gamma }{\Delta })\right] ^2
\sum_{k}\frac{(-1)^k}{\sqrt{k}}
\sin{(\frac{2\pi k\mu }{\hbar \omega _c}-\pi /4)}
\\
\label{twod}
\nonumber
\times T/\sinh{(\frac{2\pi ^2kT}{\hbar \omega _c})}
\exp{(-2\pi k\Gamma /\hbar \omega _c)}+
{\cal O}\left([max(\frac{T}{\Delta },\frac{\Gamma }{\Delta })]^4\right) ~~,
\end{eqnarray}
where we have assumed that $t>\hbar \omega _c$.
Comparing (6) and (10) we see that magnetization in a layered system
has the same form as in the homogeneous
3-D system up to the dimensionless
factor $(\sqrt{mt}\pi d)^{-1}$.
The dHvA
oscillations persist deep in the
mixed state because of the gapless
region ${\cal G}$ on the Fermi surface.
However, in the layered superconductor
such as $2H-NbSe_2$ there is a possibility of the charge density wave
instability that can modify the shape of the Fermi surface in the certain
parts of the MBZ, so that the results of the simple model
presented above might not be quantitatively appropriate.

\newpage

\begin{figure}
\caption{$\ln{[{\cal A}\sinh{(X)}H^{1/2}T^{-1}]}$ in
$V_3Si$ as function of $1/H$, where $X=2\pi ^2T/\hbar \omega _c$ and
${\cal A}={\cal A}^{\cal G}(H,T)+{\cal A}^{1-{\cal G}}(H,T)$.
Full circles represent the experimental data of
Corcoran {\it et al.} [2].  The dashed line is
a theoretical result obtained
from Eqs. (7,8), with $\Gamma$ set to $\Gamma_0=0.36$ meV,
while the full line uses
$\Gamma$ evaluated self-consistently (9).
The vertical dashed line indicates the normal-superconductor
transition at $H_{c2}=18.5 $ Tesla. We have used $\Delta_0=2.6
$ meV and the effective mass of $1.6$ $m_e$ and $1.7$ $m_e$ in the normal
and vortex state respectively. (from Ref. [3].)}
\label{fig1}
\end{figure}

\end{document}